\documentclass[12pt]{article}
\usepackage[a4paper, margin=1in]{geometry}
\usepackage{amsmath}
\usepackage{amsthm}
\usepackage{amssymb}
\usepackage{float}
\usepackage{subcaption}
\usepackage{comment}

\usepackage{mathtools}
\usepackage{appendix}
\usepackage{hyperref}

\usepackage{color}
\usepackage{graphicx}
\usepackage{bm}
\numberwithin{equation}{section}
\newtheorem{theorem}{Theorem}[section]
\newtheorem{definition}{Definition}[section]

\DeclareMathOperator{\EX}{\mathbb{E}}

\newcommand{\Q}{\mathbb{Q}}
\newcommand{\E}{\mathbb{E}}
\newcommand{\R}{\mathbb{R}}
\newcommand{\diag}{\operatorname{diag}}
\newcommand{\Var}{\operatorname{Var}}

\newcommand{\textcite}{\cite}
\newcommand{\parencite}{\cite}

\title{Pricing Generalized Dispersion Swap for Multi-Asset Stochastic Volatility Models}
\author{
	\begin{tabular}{cc}
		Semere Gebresilassie &  Mulue Gebreslasie \\
		\small School of Computing and Data Science & \small Department of Mathematics \\
		\small Wentworth Institute of Technology & \small North Dakota State University\\
		\small Email: habtemicaels@wit.edu & \small Email: mulue.gebreslasie@ndsu.edu
	\end{tabular}
	\\[2ex]
	Minglian Lin \\
	\small Department of Mathematical Sciences\\
	\small The University of Texas at El Paso \\
	\small Email: mlin2@utep.edu
}

\date{\today}

\begin{document}
	\maketitle
	\begin{abstract}
		This paper develops a generalized variance framework for pricing multi-asset dispersion swap based on the determinant of the instantaneous covariance matrix of asset log-returns. This determinant-based measure captures the joint dispersion of multivariate asset returns while accounting for their dependence structure. We formulate the proposed framework under the Heston and Barndorff-Nielsen-Shephard (BNS) stochastic volatility models, deriving an analytical pricing expression for the three-asset Heston formulation and a tractable approximation for the corresponding BNS formulation. To assess the empirical performance of the proposed approach, we conduct numerical experiments using nine stocks obtained through the quantmod package, randomly partitioned into three groups of three assets. The models are calibrated and evaluated against empirical generalized realized variance across selected time horizons.\\

	\end{abstract}
	
	\newpage
	
	\section{Introduction}
	A financial security is a financial contract whose value at maturity is determined by the price process of the underlying asset. A derivative is a financial security whose value/price is derived from one or more underlying assets. 
	There are four types of derivatives: options, forwards, futures, and swaps. A swap is a financial derivative in which two counterparties agree to exchange future cash flows, with the size of the cash flows determined at the beginning of the contract.
	Understanding the movement of a market is critical in the financial sector in order to correctly hedge and speculate on the underlying asset. Volatility and variance swaps are forward contracts on realized volatility and variance of the underlying stock, respectively. 
	In the stock market, volatility and variance are good indicators for many reasons, such as future fluctuations in stock prices. Investors or traders have insight into future fluctuations in the stock price.  
	Therefore, volatility and variance swaps provide investors with tools to hedge against or speculate on fluctuations in stock price volatility.\\
	\\
	Most of the literature has focused on pricing volatility and variance swaps for a single asset, with several studies conducted in this area under various stochastic volatility models. The classical formula, in \cite{black1973pricing}, assumed the volatility of a stock is constant. However, such an assumption for a financial model is not consistent for pricing financial derivatives in a stock market. To address this limitation, the stochastic volatility framework was introduced and has been further refined by financial researchers to reduce model risk and improve pricing accuracy. In \cite{swishchuk2004modeling}, a new probabilistic approach using the Heston model was proposed to study volatility, variance, covariance, and correlation swaps for financial markets. The impact of discrete sampling on the valuation of options on realized variance in Heston and BNS models is investigated in  \cite{ gebresilasie2024pricing, sepp2012pricing}, which establishes an analytical methodology for pricing and hedging options on realized variance in the Heston model supplemented with jumps in asset returns and variance. In \cite{marklund2015volatility}, the authors studied variance and volatility swap, and introduced the pricing strategies for some popular models (like Black-Scholes model, Merton jump diffusion model, Heston model, 3/2 model, and GARCH models). They have estimated the various parameters in the models using an option-based or price-based approach, and concluded that Black-Scholes model, Merton jump diffusion, and GARCH models do not perform well for pricing variance and volatility swaps. In some diffusion-based models, the volatility is driven by a Brownian motion, which can be correlated with the underlying asset; such models account for the different "stylized facts". Non-Gaussian processes of Ornstein-Uhlenbeck (OU) type offer the possibility of capturing important distribution deviations from Gaussianity and for flexible modeling of dependence structure \parencite{barndorff2001modelling, barndorff2001non}. The non-Gaussian OU stochastic volatility model was used by \cite{benth2007valuing} to study volatility and variance swaps. In \cite{habtemicael2016pricing}, the authors developed an analytical solution for pricing volatility and variance swaps for Barndorff-Nielsen and Shephard (BNS) process-driven financial markets. Later, in \cite{habtemicael2019volatility}, the authors modeled developed variance and volatility swap using the superposition of the BNS type model. Analytical formulas for arbitrage-free prices for the weighted variance and weighted volatility swap  are conducted by \cite{issaka2020variance} under the frame work of the BNS type stochastic volatility model.  All the aforementioned authors focused on pricing volatility and variance swaps for a single underlying asset. \\
	\\
	In today’s complex financial transactions, investors are often interested in portfolio diversification across multiple underling assets to minimize risk of an unobserved financial crisis. This motivates the development of generalized dispersion swap, which allow investors to be based on the joint behavior of two or more assets. In \cite{biswas2019proposal}, the authors proposed a framework for pricing generalized variance swaps in multi-asset financial markets with Markov-modulated volatilities. Later, \cite{biswas2020multi} they extended this framework to multi-asset financial markets driven by the BNS model. Their approach utilizes the trace and maximum Eigenvalue of the asset return covariance matrix as measures of portfolio variability. Following \cite{wilks1932certain}, the generalized variance of an \(n\)-dimensional random vector is defined as the determinant of its variance covariance matrix.  In \cite{kim2023scalar}, the authors proposed scalar measures of overall volatility and dependence for multivariate financial return series. They define the generalized variance as the determinant of the covariance matrix of the underlying asset log returns, providing a scalar measure of the overall spread of the multivariate return distribution. They also introduce the scatter coefficient, defined as the positive square root of the determinant of the correlation matrix, together with a measure of collective correlation. In \cite{hajdu2021new}, the authors  proposed the generalized variance metric to measure the composite multivariate degree of inequality (dispersion) of any multidimensional cloud. Although their work is developed in the context of multidimensional inequality and poverty, the key idea highlights the role of generalized variance as a compact measure of joint dispersion across several correlated dimensions. \\
	\\
	In this paper, we adopt the multi-asset stochastic volatility framework, while noting that alternative formulations based on Lévy-driven asset dynamics have also been proposed in the literature \parencite{lin2021analysis, lin2024estimation}.  We utilise the multi-asset stochastic volatility models under both the Heston and Barndorff-Nielsen-Shephard (BNS) frameworks. We introduce a generalized variance measure based on the covariance structure of asset returns and use it to derive a pricing approach for generalized dispersion swaps. The proposed pricing framework explicitly accounts for the units associated with a basket of assets in a portfolio. As a result, the mathematical formulation and pricing method remain closely related to established approaches for variance and volatility swaps in the literature, including \cite{swishchuk2004modeling,habtemicael2016pricing}. An important advantage of this framework is its flexibility: although it is designed for an arbitrary number of \(n\) assets, it can accommodate portfolios of different sizes and compositions without changing the underlying pricing methodology. We illustrate the approach using nine stocks organized into three portfolios of three assets each, highlighting how generalized dispersion swap prices vary across different asset combinations.\\
	\\
	The rest of the paper is organized as follows. In Section \ref{sec2}, we provide a brief overview of variance swap, generalized dispersion swap, and multi-asset return covariance matrix. In Sections \ref{sec3} and \ref{sec4}, we derive the  generalized dispersion swap for the multi-asset dynamics of the Heston and BNS model, respectively. In Section \ref{sec5}, we provide the numerical results of the model fitting and parameter estimation, Figures, tables, and the correlation matrix for generalized dispersion swap. Finally, a brief conclusion is provided in Section \ref{sec6}, and some additional derivations are included in the Appendix \ref{determinanta} and \ref{Summaryb}.

	\section{Generalized Dispersion Swap}
	\label{sec2}
	This section first reviews the conventional variance swap for a single underlying asset. We then extend the one asset realized variance framework to a multivariate system of $n$ assets using the determinant of the instantaneous covariance return matrix.
	
	For a single asset, the instantaneous variance is a scalar quantity denoted by $\sigma_t^2$. The continuously sampled realized variance is obtained by integrating
	this instantaneous variance over the life of the contract. For multiple assets, however, the instantaneous variation of the return system is described by an
	$n\times n$ covariance matrix rather than a single scalar variance.
	
	Let \(\Sigma_t\) denote the instantaneous covariance rate matrix of the log-return  of
	$n$ underlying assets. The determinant
	\(
	|\Sigma_t|
	\)
	is called the generalized variance $({GV}_t)$ of the multivariate return system. When $n=1$,
	the covariance return matrix reduces to
	\[
	\Sigma_t=\left[\sigma_t^2\right]_{1\times1}
	\]
	and hence its determinant is
	\[
	|\Sigma_t|=\sigma_t^2.
	\]
	Thus, the traditional one asset realized variance is directly extended mathematically by the generalized variance construction that was later presented. It has a distinct financial interpretation for many assets, though, as it measures the combined multivariate dispersion of the asset return system, including both individual variances and cross asset covariances, as opposed to the realized variability of a single underlying asset.\\
	
	\textit{Remark:} In our construction, we do not refer to the quantity computed from the determinant of the return covariance matrix as realized variance, since it depends on the unit of measurements. Instead, it is better understood geometrically as a measure of the hypervolume in an \(n\) dimensional space (in \cite{kim2023scalar}). This provides insight into the dispersion of a portfolio and the benefits of diversification, rather than considering a single underlying asset.
	\begin{definition}
		\label{den2.1}
		A \emph{variance swap} is a forward contract on the future realized variance of of a single underlying asset.
	\end{definition}
	The payoff at expiration \(T\) is
	\begin{equation*}
		N (\sigma_R^2(T) - K_{\text{var}}),
	\end{equation*}
	where \(N\) is the notional amount, \(\sigma_R^2\) is the realized variance, and \(K_{\text{var}}\) is the strike price.
	Calculating realized variance in discrete and continuous time have the following formulas:
	\begin{itemize}
		\item In discrete time,
		\begin{equation*}
			\sigma^{2}_{R_{d}}=\frac{m}{T(m-1)}\sum_{i=1}^{m} \left( r_i-\bar{r}\right)^2,
		\end{equation*}
		where 
		\(r_i=\log \frac{S_{i,t}}{S_{i-1,t}}
		\) is the \(i\)th logarithmic return  and \(S_{i,t}\) denotes the price of the underlying asset at time \(t_i\),  \(\bar{r}\) is the mean of the logarithmic returns and \(m\) is the number observations. Since logarithmic returns are assumed to follow a standard normal distribution, we have \(\bar{r}=0\).
		\item 
		In continuous time,
		\begin{equation*}
			\sigma^{2}_{R_{c}}=\lim_{m\to\infty} \frac{m}{T(m-1)}\sum_{i=1}^{m} \left(\log \frac{S_{i+1,t}}{S_{i,t}}\right)^2 =\frac{1}{T}\int_{0}^{T}\sigma^{2}_{t} dt,
		\end{equation*}\\
		where \(\sigma^{2}_{t}\) is the variance over an infinitesimal time period $t$.
		
		\item The above approach works for single underlying asset.  Let \(n\) denote the number of assets. For
		\(i=1,\ldots,m\), we define the vector of log returns by
		\begin{equation*}
			r_i
			=
			\begin{pmatrix}
				r_{1,i}\\
				r_{2,i}\\
				\vdots\\
				r_{n,i}
			\end{pmatrix},
			\label{return}
		\end{equation*}
		
		We define the instantaneous generalized variance as the determinant of the
		instantaneous return covariance matrix $\Sigma_t$ defined later:
		\begin{equation*}
			{GV}_{t}
			=
			\left|\Sigma_t\right|.
		\end{equation*}
		Then the realized generalized variance of the multi asset over \([0,T]\) is defined by
		\begin{equation*}
			\Sigma_{R}^2
			=
			\frac{1}{T}\int_0^T {GV}_{t}dt
			=
			\frac{1}{T}\int_0^T
			|\Sigma_t|dt.
		\end{equation*}
	\end{itemize}
	\begin{definition}
		\label{defn2.2}
		\emph{ Generalized Dispersion Swap} is a forward
		contract on the future realized generalized variance of \(n\) underlying
		assets. The realized generalized variance is constructed from the
		instantaneous covariance matrix of the asset log returns. Specifically,
		the instantaneous generalized variance is defined as the determinant of
		the instantaneous covariance matrix. This determinant provides a scalar
		measure of the joint dispersion of the multivariate return distribution,
		incorporating both the individual asset variances and their pairwise
		covariances.
		
	\end{definition}
	
	To price the determinant based generalized dispersion swap, we need to construct the return covariance matrix of the \(m\) underlying assets as given  in \eqref{return} for each $i=1,2,3,...,m$ observations.
	The portfolio return covariance matrix is obtained from the vector of individual returns that comprises variance and covariance which is given by:
	\[
	\Sigma_t=
	\begin{pmatrix}
		\text{Var}\left(r_{1,i}\right) & \text{Cov}\left(r_{1,i},r_{2,i}\right) &\cdots & \text{Cov}\left(r_{1,i},r_{n,i}\right) \\
		\text{Cov}\left(r_{2,i},r_{1,i}\right) & \text{Var}\left(r_{2,i}\right) &\cdots& \text{Cov}\left(r_{2,i},r_{n,i}\right)  \\
		\vdots & \vdots & \ddots & \vdots\\
		\text{Cov}\left(r_{n,i},r_{1,i}\right) & \text{Cov}\left(r_{n,i},r_{2,i}\right) &\cdots &  \text{Var}\left(r_{n,i}\right)
	\end{pmatrix}.
	\]
	
	The determinant based instantaneous generalized variance is given by \(|\Sigma_t|\)
	where \(\Sigma_t\) is the instantaneous return covariance matrix. Thus, the generalized dispersion swap is computed using \(|\Sigma_t|\), which measures the joint multivariate dispersion of asset returns. This approach, which is based on multivariate statistical techniques, was employed by \cite{hajdu2021new} to measure multidimensional inequality in a stratified society. The members of society form a cloud in the multidimensional space defined by inequality measures such as income, expenditure, and property. We refer the reader to \cite{hajdu2021new} for further details. A similar procedure, with additional statistical details, was also developed by \cite{kim2023scalar}. \\
	
	\noindent
	The price of a generalized dispersion swap with  strike dispersion variance price \(K_{\text{Gvar}}\) is
	\begin{equation*}
		P_{Gvar} = \EX \left[ e^{-rT} \left( \Sigma_{R}^2 - K_{\text{Gvar}} \right) \right],
	\end{equation*}
	where $\Sigma_{R}^2$ is the realized generalized variance give as:
	\begin{equation*}
		\Sigma_{R}^2 = \frac{1}{T} \int_0^T |\Sigma_t|  dt.
	\end{equation*}

	\section{Risk Neutral Multi-Asset Heston model}
	\label{sec3}

	Fix a time horizon $T>0$. Let
	\(
	(\Omega,\mathcal{F},\{\mathcal{F}_t\}_{0\leq t\leq T},\Q)
	\)
	be a filtered probability space satisfying the usual conditions. All asset price dynamics below are specified under the risk neutral pricing measure $\Q$.
	Throughout we define the following:
	\[
	X_{i,t}=\log S_{i,t},
	\quad
	i=1,\ldots,n.
	\]
	Let $ S_{1,t},\ldots,S_{n,t} $ be the prices of $n$ stocks. Under the risk-neutral measure $\Q$, we assume the stock prices satisfy
	\begin{equation*}
		\frac{d S_{i,t}}{S_{i,t}}
		=
		(r_t-q_i)d t
		+
		\sigma_{i,t}d \widetilde W_{i,t},
		\quad
		i=1,\ldots,n,
		\label{eq:heston_n_asset_price}
	\end{equation*}
	The instantaneous variance process of stock $i$ follows a Cox-Ingersoll-Ross (CIR) square root diffusion:
	\begin{equation*}
		\sigma_{i,t}^{2}
		=
		\kappa_i\left(
		\theta_i-\sigma_{i,t}^{2}
		\right) t
		+
		\xi_i\sigma_{i,t} W_{i,t},
		\quad
		\kappa_i, \theta_i, \xi_i, \sigma_{i,0}^{2}>0,
		\quad
		i=1,\ldots,n,
		\label{eq:heston_n_asset_variance}
	\end{equation*}
	where
	\begin{itemize}
		\item $\kappa_i$ is the mean-reversion speed of the variance process;
		
		\item $\theta_i$ is the long run variance level of stock $i$;
		
		\item $\xi_i$ is the volatility of variance parameter of stock $i$;
		
		\item $\sigma_{i,0}^{2}$ is the initial instantaneous variance of stock $i$.
	\end{itemize}

	{Brownian motion correlation structure.}
	Let
	
	$$
	\widetilde{\mathbf W}_t
	=
	(\widetilde W_{1,t},\ldots,\widetilde W_{n,t})^\top,
	\qquad
	\mathbf W_t
	=
	(W_{1,t},\ldots,W_{n,t})^\top,
	$$
	
	denote the Brownian motions driving the stock prices and their corresponding variance processes, respectively. Their instantaneous correlations are specified by
	
	$$
	d\langle \widetilde W_i,\widetilde W_j\rangle_t
	=c_{ij}\,dt,\qquad
	d\langle W_i,W_j\rangle_t
	=c_{ij}^{*}\,dt,
	$$
	
	and
	
	$$
	d\langle \widetilde W_i,W_j\rangle_t
	=c_{ij}^{**}\,dt,
	\qquad i,j=1,\ldots,n.
	$$
	
	Here,
	
	$$
	C=(c_{ij}),\qquad
	C^{*}=(c_{ij}^{*}),\qquad
	C^{**}=(c_{ij}^{**}),
	$$
	
	where \(C\) and \(C^{*}\) are symmetric positive-semidefinite correlation matrices with \(c_{ii}=c_{ii}^{*}=1\). The diagonal elements of \(C^{**}\),
	
	$$
	c_{ii}^{**}=\eta_i,\qquad -1\leq\eta_i\leq1,
	$$
	
	represent the leverage correlation between the stock price and its own variance process:
	
	$$
	d\langle\widetilde W_i,W_i\rangle_t=\eta_i\,dt.
	$$
	
	Equivalently, the \(2n\) Brownian motions are jointly characterized by the block correlation matrix
	
	$$
	\boldsymbol{\Sigma}
	=
	\begin{pmatrix}
		C & C^{**}\\
		(C^{**})^\top & C^{*}
	\end{pmatrix},
	\qquad
	\boldsymbol{\Sigma}\succeq0,
	$$
	
	which guarantees that
	\((\widetilde{\mathbf W}_t^\top,\mathbf W_t^\top)^\top\)
	is well defined on a common filtered probability space.

	Applying It\^{o}'s formula to
	\(
	X_{i,t}=\log S_{i,t}
	\)
	gives
	\begin{equation*}
		dX_{i,t}
		=
		\left(
		r_t-q_i-\frac12\sigma_{i,t}^{2}
		\right) dt
		+
		\sigma_{i,t} d\widetilde W_{i,t},
		\quad
		i=1,\ldots,n.
		\label{eq:heston_n_asset_log_price}
	\end{equation*}
	The continuous quadratic covariation between the log-prices of the $i$th and $j$th assets
	is
	\begin{equation*}
		d\langle X_i,X_j\rangle_t
		=
		c_{ij}
		\sigma_{i,t}\sigma_{j,t}
		dt.
		\label{eq:heston_n_asset_quadratic_covariation}
	\end{equation*}

	For each asset $i$, the risk-neutral conditional mean of the CIR variance process is
	\begin{equation*}
		\E^\Q\left[
		\sigma_{i,t}^{2}
		\mid
		\mathcal{F}_0
		\right]
		=
		\theta_i
		+
		\left(
		\sigma_{i,0}^{2}-\theta_i
		\right)
		e^{-\kappa_i t}.
		\label{eq:heston_n_asset_variance_mean}
	\end{equation*}

	\paragraph{Illustration with a Three Stock Heston Model:}
	We consider the general Heston model to three stocks $S_{1,t}$, $S_{2,t}$, and $S_{3,t}$.
	Under $\Q$, the risk-neutral stock price dynamics are
	\begin{align*}
		\frac{ dS_{1,t}}{S_{1,t}}
		&=
		(r_t-q_1) t
		+
		\sigma_{1,t} d\widetilde W_{1,t},
		\\
		\frac{d S_{2,t}}{S_{2,t}}
		&=
		(r_t-q_2) t
		+
		\sigma_{2,t} d\widetilde W_{2,t},
		\\
		\frac{ dS_{3,t}}{S_{3,t}}
		&=
		(r_t-q_3) t
		+
		\sigma_{3,t} d\widetilde W_{3,t},
	\end{align*}
	where the three variance processes satisfy
	\begin{align*}
		d \sigma_{1,t}^{2}
		&=
		\kappa_1
		\left(
		\theta_1-\sigma_{1,t}^{2}
		\right) t
		+
		\xi_1\sigma_{1,t} dW_{1,t},
		\quad
		\sigma_{1,0}^{2}>0,
		\\
		d \sigma_{2,t}^{2}
		&=
		\kappa_2
		\left(
		\theta_2-\sigma_{2,t}^{2}
		\right) t
		+
		\xi_2\sigma_{2,t} dW_{2,t},
		\quad
		\sigma_{2,0}^{2}>0,
		\\
		d \sigma_{3,t}^{2}
		&=
		\kappa_3
		\left(
		\theta_3-\sigma_{3,t}^{2}
		\right) t
		+
		\xi_3\sigma_{3,t} dW_{3,t},
		\quad
		\sigma_{3,0}^{2}>0.
	\end{align*}
	Then, the stock return correlation matrix is
	\begin{equation*}
		C
		=
		\begin{pmatrix}
			1 & c_{12} & c_{13}\\
			c_{12} & 1 & c_{23}\\
			c_{13} & c_{23} & 1
		\end{pmatrix}.
		\label{eq:heston_three_stock_return_correlation_matrix}
	\end{equation*}
	Equivalently,
	\begin{align*}
		d\langle \widetilde W_1,\widetilde W_2\rangle_t
		=
		c_{12} dt,
		\quad
		d\langle \widetilde W_1,\widetilde W_3\rangle_t
		=
		c_{13} dt,
		\quad \text{and} \quad
		d\langle \widetilde W_2,\widetilde W_3\rangle_t
		=
		c_{23} dt.
	\end{align*}

	In this work, we assume independent variance factors and zero cross asset price variance correlations:
	\begin{align*}
		d\langle W_i,W_j\rangle_t = 0
		\quad
		\text{and}
		\quad
		d\langle\widetilde W_i,W_j\rangle_t = 0,
		\quad
		\forall i\neq j.
	\end{align*}
	The corresponding log-price dynamics are
	\begin{align*}
		dX_{1,t}
		&=
		\left(
		r_t-q_1-\frac12\sigma_{1,t}^{2}
		\right) t
		+
		\sigma_{1,t} d\widetilde W_{1,t},
		\\
		dX_{2,t}
		&=
		\left(
		r_t-q_2-\frac12\sigma_{2,t}^{2}
		\right)t
		+
		\sigma_{2,t} d\widetilde W_{2,t},
		\\
		dX_{3,t}
		&=
		\left(
		r_t-q_3-\frac12\sigma_{3,t}^{2}
		\right) t
		+
		\sigma_{3,t} d\widetilde W_{3,t}.
	\end{align*}
	Hence, the $3\times3$ instantaneous covariance matrix of the continuous log-return components is
	\begin{equation}
		\Sigma_t^{H}
		=
		\begin{pmatrix}
			\sigma_{1,t}^{2}
			&
			c_{12}\sigma_{1,t}\sigma_{2,t}
			&
			c_{13}\sigma_{1,t}\sigma_{3,t}
			\\[4pt]
			c_{12}\sigma_{1,t}\sigma_{2,t}
			&
			\sigma_{2,t}^{2}
			&
			c_{23}\sigma_{2,t}\sigma_{3,t}
			\\[4pt]
			c_{13}\sigma_{1,t}\sigma_{3,t}
			&
			c_{23}\sigma_{2,t}\sigma_{3,t}
			&
			\sigma_{3,t}^{2}
		\end{pmatrix}.
		\label{eq:heston_three_covariance_matrix}
	\end{equation}
	Furthermore, we define
	\[
	D_t^{H}
	=
	\diag
	\left(
	\sigma_{1,t},
	\sigma_{2,t},
	\sigma_{3,t}
	\right).
	\]
	Then,
	\begin{equation*}
		\Sigma_t^{H}
		=
		D_t^{H}CD_t^{H}.
		\label{eq:heston_three_drd}
	\end{equation*}
	Now, we need to compute \(\sigma^2_{R}\), where \(\sigma^2_{R}\) denotes the realized variance computed from the determinant of the portfolio covariance matrix \(\Sigma_t\). It is defined as the determinant based instantaneous generalized variance  in the period [0,T], which is given by
	\begin{equation}
		\sigma^2_{R} = \frac{1}{T}\int^T_0 \left|\Sigma^H_t\right|dt.
		\label{eq:hes20}
	\end{equation}
	The determinant of \eqref{eq:heston_three_covariance_matrix} is given by the product of the squared volatilities and the determinant of the correlation coefficient matrix, as shown below. For a detailed proof of the following equation, we refer the reader to Appendix \ref{determinanta}.
	\begin{align}
		\left| \Sigma^H_t \right| = |C| \prod_{i=1}^{3} \sigma_{i,t}^{2},
		\label{eq:hes21}
	\end{align}
	where $ C = (c_{ij})_{1\leq i,j \leq 3} $ is the correlation matrix of stock prices calculated by the stock price data.
	By the independence of $\sigma_{i,t}^{2}$, $ i = 1, 2, 3 $, we have
	\begin{align}
		\EX^\Q \left[ \left| \Sigma^H_t \right| \right] = |C| \prod_{i=1}^{3} \EX^\Q [\sigma_{i,t}^{2}].
		\label{eq:hes22}
	\end{align}
	Following \eqref{eq:heston_n_asset_variance_mean}, we obtain
	\begin{align}
		& \prod_{i=1}^{3} \EX^\Q [\sigma_{i,t}^{2}] \nonumber\\
		= \ 
		& e^{-(k_3+k_2+k_1) t} \big(\sigma_{3,0}^{2} - \theta_3\big) \big(\sigma_{2,0}^{2} - \theta_2\big) \big(\sigma_{1,0}^{2} - \theta_1\big)
		+ e^{-(k_3+k_2) t} \big(\sigma_{3,0}^{2} - \theta_3\big) \big(\sigma_{2,0}^{2} - \theta_2\big) \theta_1 \notag\\
		& + e^{-(k_3+k_1) t} \big(\sigma_{3,0}^{2} - \theta_3\big) \big(\sigma_{1,0}^{2} - \theta_1\big) \theta_2
		+ e^{-(k_2+k_1) t} \big(\sigma_{2,0}^{2} - \theta_2\big) \big(\sigma_{1,0}^{2} - \theta_1\big) \theta_3 \notag\\
		& + e^{-k_3 t} \big(\sigma_{3,0}^{2} - \theta_3\big) \theta_2 \theta_1
		+ e^{-k_2 t} \big(\sigma_{2,0}^{2} - \theta_2\big) \theta_3 \theta_1
		+ e^{-k_1 t} \big(\sigma_{1,0}^{2} - \theta_1\big) \theta_3 \theta_2
		+ \theta_3 \theta_2 \theta_1.
		\label{eq:hes23}
	\end{align}
	
	For an arbitrary number \(n\) of underlying assets, equation (\ref{eq:hes21}) can be generalized as follows 
	\begin{align*}
		\left| \Sigma^H_t \right| = |C| \prod_{i=1}^{n} \sigma_{i,t}^{2}.
	\end{align*}
	
	However, the mathematical expression for the expected value becomes considerably lengthy and complex as the dimension increases. Therefore, to maintain clarity and readability, the explicit formulation of the expected value is omitted.
	
	\begin{theorem}
		The arbitrage free price of the  generalized dispersion swap for \(S_{1,t}, S_{2,t},\) and \(S_{3,t}\), under the Heston model and using the generalized variance method, is given by:
		\begin{equation*}
			P_{Gvar} = e^{-rT} \EX^\Q \left[\sigma^2_{R}\right] - e^{-rT} K_{Gvar}.
		\end{equation*}
	\end{theorem}
	\begin{proof} 
		The expected value of equation \eqref{eq:hes20} applying equations \eqref{eq:hes22} and \eqref{eq:hes23} gives that 
		\begin{align}
			\EX^\Q \left[ \sigma^2_{R} \right] 
			& = \frac{1}{T} \int^T_0 \EX^\Q \left[ \left| \Sigma^H_t \right| \right] dt \notag\\
			& = \frac{|C|}{T}
			\bigg[
			\Big( \frac{1-e^{-(k_3+k_2+k_1)T}}{k_3+k_2+k_1}\Big)
			\big(\sigma_{3,0}^{2} - \theta_3\big) \big(\sigma_{2,0}^{2} - \theta_2\big) \big(\sigma_{1,0}^{2} - \theta_1\big) \notag\\
			&\qquad \quad
			+ \Big(\frac{1-e^{-(k_3+k_2)T}}{k_3+k_2}\Big)
			\big(\sigma_{3,0}^{2} - \theta_3\big) \big(\sigma_{2,0}^{2} - \theta_2\big) \theta_1 \notag\\
			&\qquad \quad
			+ \Big( \frac{1-e^{-(k_3+k_1)T}}{k_3+k_1}\Big)
			\big(\sigma_{3,0}^{2} - \theta_3\big) \big(\sigma_{1,0}^{2} - \theta_1\big) \theta_2 \notag\\
			&\qquad \quad
			+ \Big( \frac{1-e^{-(k_2+k_1)T}}{k_2+k_1}\Big)
			\big(\sigma_{2,0}^{2} - \theta_2\big) \big(\sigma_{1,0}^{2} - \theta_1\big) \theta_3 \notag\\
			&\qquad \quad
			+ \Big(\frac{1-e^{-k_3T}}{k_3}\Big) \big(\sigma_{3,0}^{2} - \theta_3\big) \theta_2 \theta_1
			+ \Big( \frac{1-e^{-k_2T}}{k_2}\Big) \big(\sigma_{2,0}^{2} - \theta_2\big) \theta_3 \theta_1 \notag\\
			&\qquad \quad
			+ \Big( \frac{1-e^{-k_1T}}{k_1}\Big) \big(\sigma_{1,0}^{2} - \theta_1\big) \theta_3 \theta_2
			+ T \theta_3 \theta_2 \theta_1
			\bigg].
			\label{eq:hes25}
		\end{align}
		So the theorem uses equation~\eqref{eq:hes25} to complete the proof.
	\end{proof}

	\section{Risk-Neutral Multi-Asset BNS Model}
	\label{sec4}
	Let $ S_{1,t},\ldots,S_{n,t} $ be $n$ stock prices and let $ X_{i,t}=\log S_{i,t} $.
	Under the risk-neutral measure $\Q$, we consider the multi-asset BNS model
	\begin{equation*}
		d X_{i,t}
		=
		b_{i,t}d t
		+
		\sigma_{i,t}d W_{i,t}
		+
		\rho_id Z_{\lambda t}^{(0)},
		\quad
		i=1,\ldots,n,
		\label{eq:bns_n_asset_log_price}
	\end{equation*}
	where
	\begin{itemize}
		\item $W_{i,t}$ is the Brownian motion driving the continuous return component
		of stock $i$;
		
		\item $\sigma_{i,t}^{2}$ is the instantaneous variance of stock $i$;
		
		\item $Z^{(0)}$ is a common subordinator generating common jump risk;
		
		\item $\lambda>0$ is the time-scaling parameter of the common subordinator;
		
		\item $\rho_i\in\R$ is the loading of asset $i$ on the common jump factor;
		
		\item $b_{i,t}$ is the risk-neutral log-price drift.
	\end{itemize}
	
	Let the Brownian motions satisfy
	\begin{equation*}
		d\langle W_i,W_j\rangle_t
		=
		c_{ij}^{B}d t,
		\quad
		i,j=1,\ldots,n,
		\label{eq:bns_n_asset_brownian_correlation}
	\end{equation*}
	where $ R^{B}
	=
	\left(
	c_{ij}^{B}
	\right)_{1\leq i,j\leq n}$
	is a symmetric positive-semidefinite correlation matrix with $ c_{ii}^{B}=1 $.
	For each asset $i$, we let the squared volatility follows an Ornstein-Uhlenbeck type process driven
	by a subordinator:
	\begin{equation}
		d\sigma_{i,t}^{2}
		=
		-\lambda_i\sigma_{i,t}^{2}d t
		+
		d Z_{\lambda_i t}^{(i)},
		\quad
		\lambda_i, \sigma_{i,0}^{2}>0,
		\quad
		i=1,\ldots,n.
		\label{eq:bns_n_asset_variance}
	\end{equation}
	The subordinators $ Z^{(0)},Z^{(1)},\ldots,Z^{(n)} $ are assumed mutually independent and independent of the Brownian motion vector $ (W_1,\ldots,W_n) $.

	We define the first and second cumulants by
	\[
	\kappa_{1,i}
	=
	\E^\Q \left[
	Z_1^{(i)}
	\right]
	\quad
	\text{and}
	\quad
	\kappa_{2,i}
	=
	\Var^\Q \left[
	Z_1^{(i)}
	\right].
	\]
	Then,
	\[
	\kappa_{1,0}
	=
	\E^\Q \left[
	Z_1^{(0)}
	\right]
	\quad
	\text{and}
	\quad
	\kappa_{2,0}
	=
	\Var^\Q \left[
	Z_1^{(0)}
	\right].
	\]

	Since $ S_{i,t}=S_{i,0} e^{X_{i,t}} $, the drift $b_{i,t}$ must be chosen so that the discounted cum-dividend asset price
	is a $\Q$-martingale. 
	A sufficient risk-neutral drift specification is
	\begin{equation*}
		b_{i,t}
		=
		r_t-q_i
		-\frac12\sigma_{i,t}^{2}
		-\lambda\kappa_0(\rho_i).
		\label{eq:bns_n_asset_risk_neutral_drift}
	\end{equation*}
	Hence, the risk-neutral BNS log-price dynamics can be written as
	\begin{equation}
		d X_t^i
		=
		\left[
		r_t-q_i
		-\frac12\sigma_{i,t}^{2}
		-\lambda_0\kappa_0(\rho_i)
		\right]d t
		+
		\sigma_{i,t}d W_{i,t}
		+
		\rho_id Z_{\lambda t}^{(0)}.
		\label{eq:bns_n_asset_risk_neutral_log_price}
	\end{equation}
	It is noteworthy that the compensator term $ -\lambda_0\kappa_0(\rho_i) $ ensures that the jump component does not introduce an arbitrage opportunity under the pricing measure.

	The solution to \eqref{eq:bns_n_asset_variance} is
	\begin{equation*}
		\sigma_{i,t}^{2}
		=
		e^{-\lambda_i t}\sigma_{i,0}^{2}
		+
		\int_0^t
		e^{-\lambda_i(t-s)}
		d Z_{\lambda_i s}^{(i)}.
		\label{eq:bns_n_asset_variance_solution}
	\end{equation*}
	
	Since $Z^{(i)}$ is a subordinator and $\sigma_{i,0}^{2}>0$, we have
	\begin{equation*}
		\sigma_{i,t}^{2}
		\geq
		e^{-\lambda_i t}\sigma_{i,0}^{2}
		>
		0,
		\quad
		0\leq t\leq T.
		\label{eq:bns_n_asset_variance_positivity}
	\end{equation*}
	Hence, the risk-neutral first moment is
	\begin{equation*}
		\E^\Q \left[
		\sigma_{i,t}^{2}
		\right]
		=
		e^{-\lambda_i t}\sigma_{i,0}^{2}
		+
		\kappa_{1,i}
		\left(
		1-e^{-\lambda_i t}
		\right).
		\label{eq:bns_n_asset_variance_mean}
	\end{equation*}

	The continuous local-martingale part of \eqref{eq:bns_n_asset_risk_neutral_log_price} is $ \sigma_{i,t}d W_{i,t} $.
	Therefore, the continuous covariance-rate matrix has entries
	\begin{equation*}
		\left(
		\Sigma_t^{B,c}
		\right)_{ij}
		=
		c_{ij}^{B}
		\sigma_{i,t}\sigma_{j,t}.
		\label{eq:bns_n_asset_continuous_covariance_entries}
	\end{equation*}
	We define
	\[
	D_t^B
	=
	\diag
	\left(
	\sigma_{1,t},
	\ldots,
	\sigma_{n,t}
	\right).
	\]
	Then,
	\begin{equation*}
		\Sigma_t^{B,c}
		=
		D_t^B R^B D_t^B.
		\label{eq:bns_n_asset_continuous_covariance}
	\end{equation*}
	
	If one wishes to include the predictable covariance contribution from the common
	jump factor, we define
	\[
	{\rho}
	=
	\begin{pmatrix}
		\rho_1\\
		\vdots\\
		\rho_n
	\end{pmatrix}.
	\]
	Then the instantaneous second-moment-rate matrix is
	\begin{equation*}
		\widetilde{\Sigma}_t^B
		=
		D_t^B R^B D_t^B
		+
		\lambda\kappa_{2,0}
		{\rho}{\rho}^{\top}.
		\label{eq:bns_n_asset_jump_inclusive_covariance}
	\end{equation*}
	It is noteworthy that the matrices $\Sigma_t^{B,c}$ and $\widetilde{\Sigma}_t^B$ represent different
	objects. 
	The first is the covariance-rate matrix of the continuous local-martingale component. 
	The second is a predictable second-moment-rate matrix that also includes the covariance contribution of the common jump factor.

	\paragraph{Illustration with a Three Stock BNS Model:}
	
	We consider the BNS model to three stocks $ S_{1,t}$, $S_{2,t}$, and $S_{3,t}$.
	The risk-neutral log-price processes are
	\begin{align*}
		d X_{1,t}
		&=
		\left[
		r_t-q_1
		-\frac12\sigma_{1,t}^{2}
		-\lambda_0\kappa_0(\rho_1)
		\right]d t
		+
		\sigma_{1,t}d W_{1,t}
		+
		\rho_1d Z_{\lambda t}^{(0)},
		\\
		d X_{2,t}
		&=
		\left[
		r_t-q_2
		-\frac12\sigma_{2,t}^{2}
		-\lambda_0\kappa_0(\rho_2)
		\right]d t
		+
		\sigma_{2,t}d W_{2,t}
		+
		\rho_2d Z_{\lambda t}^{(0)},
		\\
		d X_{3,t}
		&=
		\left[
		r_t-q_3
		-\frac12\sigma_{3,t}^{2}
		-\lambda_0 \kappa_0(\rho_3)
		\right]d t
		+
		\sigma_{3,t}d W_{3,t}
		+
		\rho_3d Z_{\lambda t}^{(0)},
	\end{align*}
	where the three squared-volatility processes satisfy
	\begin{align*}
		d \sigma_{1,t}^{2}
		&=
		-\lambda_1\sigma_{1,t}^{2}d t
		+
		d Z_{\lambda_1t}^{(1)},
		\quad
		\sigma_{1,0}^{2}>0,
		\\
		d \sigma_{2,t}^{2}
		&=
		-\lambda_2\sigma_{2,t}^{2}d t
		+
		d Z_{\lambda_2t}^{(2)},
		\quad
		\sigma_{2,0}^{2}>0,
		\\
		d \sigma_{3,t}^{2}
		&=
		-\lambda_3\sigma_{3,t}^{2}d t
		+
		d Z_{\lambda_3t}^{(3)},
		\quad
		\sigma_{3,0}^{2}>0.
	\end{align*}
	Then, the Brownian return-correlation matrix is
	\begin{equation*}
		R^B
		=
		\begin{pmatrix}
			1 & \rho_{12}^{B} & \rho_{13}^{B}\\
			\rho_{12}^{B} & 1 & \rho_{23}^{B}\\
			\rho_{13}^{B} & \rho_{23}^{B} & 1
		\end{pmatrix}.
		\label{eq:bns_three_stock_brownian_correlation}
	\end{equation*}
	
	Equivalently,
	\begin{align*}
		d\langle W_1,W_2\rangle_t = \rho_{12}^{B}d t,
		\quad
		d\langle W_1,W_3\rangle_t = \rho_{13}^{B}d t,
		\quad \text{and} \quad
		d\langle W_2,W_3\rangle_t = \rho_{23}^{B}d t.
	\end{align*}
	Hence, the continuous covariance-rate matrix of the three log-return processes is
	\begin{equation}
		\Sigma_t^{B,c}
		=
		\begin{pmatrix}
			\sigma_{1,t}^{2}
			&
			\rho_{12}^{B}\sigma_{1,t}\sigma_{2,t}
			&
			\rho_{13}^{B}\sigma_{1,t}\sigma_{3,t}
			\\[4pt]
			\rho_{12}^{B}\sigma_{1,t}\sigma_{2,t}
			&
			\sigma_{2,t}^{2}
			&
			\rho_{23}^{B}\sigma_{2,t}\sigma_{3,t}
			\\[4pt]
			\rho_{13}^{B}\sigma_{1,t}\sigma_{3,t}
			&
			\rho_{23}^{B}\sigma_{2,t}\sigma_{3,t}
			&
			\sigma_{3,t}^{2}
		\end{pmatrix}.
		\label{eq:bns_three_stock_continuous_covariance}
	\end{equation}
	Furthermore, we define
	\[
	D_t^B
	=
	\diag
	\left(
	\sigma_{1,t},
	\sigma_{2,t},
	\sigma_{3,t}
	\right).
	\]
	Then,
	\begin{equation*}
		\Sigma_t^{B,c}
		=
		D_t^B R^B D_t^B.
		\label{eq:bns_three_stock_drd}
	\end{equation*}
	
	For common jump factor,we define
	\[
	{\rho}
	=
	\begin{pmatrix}
		\rho_1\\
		\rho_2\\
		\rho_3
	\end{pmatrix},
	\quad
	\kappa_{2,0}
	=
	\Var^\Q\left[Z_1^{(0)}\right].
	\]
	Hence, the jump-inclusive predictable second-moment-rate matrix is
	\begin{equation*}
		\widetilde{\Sigma}_t^B
		=
		\Sigma_t^{B,c}
		+
		\lambda \kappa_{2,0}
		{\rho}{\rho}^{\top}.
		\label{eq:bns_three_stock_jump_inclusive_covariance}
	\end{equation*}
	
	Equivalently,
	\begin{equation*}
		\widetilde{\Sigma}_t^B
		=
		\begin{pmatrix}
			\sigma_{1,t}^{2}
			+
			\lambda \kappa_{2,0}\rho_1^2
			&
			\rho_{12}^{B}\sigma_{1,t}\sigma_{2,t}
			+
			\lambda \kappa_{2,0}\rho_1\rho_2
			&
			\rho_{13}^{B}\sigma_{1,t}\sigma_{3,t}
			+
			\lambda \kappa_{2,0}\rho_1\rho_3
			\\[4pt]
			\rho_{12}^{B}\sigma_{1,t}\sigma_{2,t}
			+
			\lambda \kappa_{2,0}\rho_1\rho_2
			&
			\sigma_{2,t}^{2}
			+
			\lambda \kappa_{2,0}\rho_2^2
			&
			\rho_{23}^{B}\sigma_{2,t}\sigma_{3,t}
			+
			\lambda \kappa_{2,0}\rho_2\rho_3
			\\[4pt]
			\rho_{13}^{B}\sigma_{1,t}\sigma_{3,t}
			+
			\lambda \kappa_{2,0}\rho_1\rho_3
			&
			\rho_{23}^{B}\sigma_{2,t}\sigma_{3,t}
			+
			\lambda \kappa_{2,0}\rho_2\rho_3
			&
			\sigma_{3,t}^{2}
			+
			\lambda \kappa_{2,0}\rho_3^2
		\end{pmatrix}.
		\label{eq:bns_three_stock_jump_inclusive_full_matrix}
	\end{equation*}

	Similar to equation \eqref{eq:hes20}, we also need to compute the realized variance $\sigma^2_{R}$ over the period $[0, T]$, which is given by
	\begin{align} \label{sigma^2_R}
		\sigma_{R}^2 = \frac{1}{T}\int^T_0 \left|\Sigma_t^B\right| dt.	
	\end{align}
	In Appendix \ref{determinanta}, we obtain 
	\begin{align} \label{Sigma_2_det}
		& \left| \Sigma_t^B \right| = \nonumber\\ 
		& |C| \prod_{i=1}^{3}\sigma_{i,t}^2 
		+ \lambda  \kappa_{2,0} |C| \Big(
		\delta_{11}\rho_1^2 \sigma_{3,t}^2 \sigma_{2,t}^2 
		+ \delta_{22}\rho_2^2 \sigma_{3,t}^2 \sigma_{1,t}^2 
		+ \delta_{33}\rho_3^2 \sigma_{2,t}^2 \sigma_{1,t}^2 \nonumber\\
		& \qquad \qquad \qquad \qquad \qquad
		+ 2 \delta_{21}\rho_2\rho_1 \sigma_{3,t}^2 \sigma_{2,t} \sigma_{1,t}
		+ 2 \delta_{31}\rho_3\rho_1 \sigma_{3,t} \sigma_{2,t}^2 \sigma_{1,t} 
		+ 2 \delta_{32}\rho_3\rho_2 \sigma_{3,t} \sigma_{2,t} \sigma_{1,t}^2 \Big).
	\end{align}
	We then have
	\begin{align} \label{E_Sigma_2}
		& \EX^\Q \left[\left| \Sigma_t^B \right|\right] = \nonumber\\
		& |C| \prod_{i=1}^{3}\EX^\Q [\sigma_{i,t}^2]
		+ \lambda  \kappa_{2,0} |C| \Big(
		\delta_{11}\rho_1^2 \EX^\Q[\sigma_{3,t}^2] \EX^\Q[\sigma_{2,t}^2] 
		+ \delta_{22}\rho_2^2 \EX^\Q[\sigma_{3,t}^2] \EX^\Q[\sigma_{1,t}^2] \nonumber\\
		& \qquad \qquad \qquad \qquad \
		+ \delta_{33}\rho_3^2 \EX^\Q[\sigma_{2,t}^2] \EX^\Q[\sigma_{1,t}^2]
		+ 2 \delta_{21}\rho_2\rho_1 \EX^\Q[\sigma_{3,t}^2] \EX^\Q[\sigma_{2,t}] \EX^\Q[\sigma_{1,t}] \nonumber\\
		& \qquad \qquad \qquad \qquad \
		+ 2 \delta_{31}\rho_3\rho_1 \EX^\Q[\sigma_{3,t}] \EX^\Q[\sigma_{2,t}^2] \EX^\Q[\sigma_{1,t}]
		+ 2 \delta_{32}\rho_3\rho_2 \EX^\Q[\sigma_{3,t}] \EX^\Q[\sigma_{2,t}] \EX^\Q[\sigma_{1,t}^2] \Big).
	\end{align}
	Next, we compute the expected value of the variance process:
	\begin{align} \label{E}
		\EX^\Q[\sigma_{2,t}] 
		& = e^{-\lambda t} \EX^\Q [\sigma_{2,0}] + \int_0^t e^{-\lambda (t-s)} \EX^\Q [dZ_{\lambda s}] \nonumber\\
		& = e^{-\lambda t} \sigma_{2,0} + \int_0^t e^{-\lambda (t-s)} \ \kappa_1 \lambda ds \nonumber\\
		& = e^{-\lambda t}(\sigma_{2,0} - \kappa_1) + \kappa_1,
	\end{align}
	where $ \kappa_1 $ is the 1st cumulant of the subordinator driving the variance process, i.e. the mean of jump size, which can be calculated by the stock price data.
	In the 3-dimensional case, we have
	\begin{align*}
		\EX^\Q[\sigma_{i,t}^2] = e^{-\lambda t}(\sigma_{i,0}^2 - \kappa_{i,1}) + \kappa_{i,1}, \quad i = 1,2,3.
	\end{align*}
	Using the above equation, we obtain the followings:
	\begin{align} \label{E0}
		& \prod_{i=1}^{3} \mathbb{E}^\Q[\sigma_{i,t}^2] \nonumber\\
		=\ & 
		e^{-3\lambda t} (\sigma_{1,0}^2 - \kappa_{1,1})(\sigma_{2,0}^2 - \kappa_{2,1})(\sigma_{3,0}^2 - \kappa_{3,1}) \nonumber\\
		& 
		+ e^{-2\lambda t} \Big[
		\kappa_{1,1} (\sigma_{2,0}^2 - \kappa_{2,1})(\sigma_{3,0}^2 - \kappa_{3,1})
		+ \kappa_{2,1} (\sigma_{1,0}^2 - \kappa_{1,1})(\sigma_{3,0}^2 - \kappa_{3,1}) \nonumber\\
		&\qquad \qquad 
		+ \kappa_{3,1} (\sigma_{1,0}^2 - \kappa_{1,1})(\sigma_{2,0}^2 - \kappa_{2,1})
		\Big] \nonumber\\
		& 
		+ e^{-\lambda t} \Big[
		\kappa_{1,1} \kappa_{2,1} (\sigma_{3,0}^2 - \kappa_{3,1}) +
		\kappa_{1,1} \kappa_{3,1} (\sigma_{2,0}^2 - \kappa_{2,1}) +
		\kappa_{2,1} \kappa_{3,1} (\sigma_{1,0}^2 - \kappa_{1,1})
		\Big]
		+ \kappa_{1,1} \kappa_{2,1} \kappa_{3,1},
	\end{align}
	
	\begin{align} \label{E1}
		& \mathbb{E}^\Q [\sigma_{3,t}^2] \mathbb{E}^\Q [\sigma_{2,t}^2] \nonumber\\
		=\ & e^{-2\lambda t} (\sigma_{3,0}^2 - \kappa_{3,1})(\sigma_{2,0}^2 - \kappa_{2,1})
		+ e^{-\lambda t} \Big[ 
		\kappa_{3,1}(\sigma_{2,0}^2 - \kappa_{2,1})
		+ \kappa_{2,1}(\sigma_{3,0}^2 - \kappa_{3,1})
		\Big] 
		+ \kappa_{3,1} \kappa_{2,1},
	\end{align}
	
	\begin{align} \label{E2}
		& \mathbb{E}^\Q [\sigma_{3,t}^2] \mathbb{E}^\Q [\sigma_{1,t}^2] \nonumber\\
		=\ & e^{-2\lambda t} (\sigma_{3,0}^2 - \kappa_{3,1})(\sigma_{1,0}^2 - \kappa_{1,1})
		+ e^{-\lambda t} \Big[
		\kappa_{3,1}(\sigma_{1,0}^2 - \kappa_{1,1})
		+ \kappa_{1,1}(\sigma_{3,0}^2 - \kappa_{3,1})
		\Big] 
		+ \kappa_{3,1} \kappa_{1,1},
	\end{align}
	
	\begin{align} \label{E3}
		& \mathbb{E}^\Q [\sigma_{2,t}^2] \mathbb{E}^\Q [\sigma_{1,t}^2] \nonumber\\
		=\ & e^{-2\lambda t} (\sigma_{2,0}^2 - \kappa_{2,1})(\sigma_{1,0}^2 - \kappa_{1,1})
		+ e^{-\lambda t} \Big[
		\kappa_{2,1}(\sigma_{1,0}^2 - \kappa_{1,1})
		+ \kappa_{1,1}(\sigma_{2,0}^2 - \kappa_{2,1})
		\Big]
		+ \kappa_{2,1} \kappa_{1,1}.
	\end{align}
	
	\begin{theorem}
		The arbitrage free price of the generalized dispersion swap for \(S_{1,t}, S_{2,t},\) and \(S_{3,t}\), assuming they follow the BNS model and using the generalized variance method, is given by:	
		\begin{equation*}
			P_{Gvar} = e^{-rT} \EX^\Q \left[\sigma_{R}^2\right] - e^{-rT} K_{\text{Gvar}}.
		\end{equation*}			
	\end{theorem}
	\begin{proof} 
		We firstly compute 
		\begin{align} \label{Var}
			\text{Var}^\Q[\sigma_{2,t}] 
			& = 0 + \int_0^t \big(e^{-\lambda (t-s)}\big)^2\ \text{Var}^\Q[dZ_{\lambda s}] + 0 \nonumber\\
			& = \int_0^t e^{-2\lambda (t-s)}\ \kappa_2 \lambda ds \nonumber\\
			& = \frac{\kappa_2}{2}\left(1 - e^{-2\lambda t}\right),
		\end{align}
		where $ \kappa_2 $ is the 2nd cumulant of the subordinator driving the variance process, i.e. the variance of jump size, which can be also calculated by the stock price data.
		Then, a useful estimate approximation regarding the expected value of realized volatility is obtained in \cite{brockhaus2000volatility} and is given by
		\begin{align}\label{Approx}
			\mathbb{E}^\Q[\sigma_t] = \EX^\Q \left[\sqrt{\sigma_{2,t}}\right] 
			\approx \sqrt{\EX^\Q[\sigma_{2,t}]} - \frac{\text{Var}^\Q[\sigma_{2,t}]}{8(\EX^\Q[\sigma_{2,t}])^{3/2}}. 
		\end{align}
		The absolute error of such approximation is less than or equal to $ \displaystyle \frac{\mu_3}{16(\EX^\Q[\sigma_{2,t}])^{5/2}}$, where $\mu_3$ is the 3rd central moment of $ \sigma_{2,t} $.
		In the 3-dimensional case, we substitute \eqref{E} and \eqref{Var} into \eqref{Approx} and obtain
		\begin{align} \label{Var_i}
			\mathbb{E}^\Q[\sigma_{i,t}] 
			\approx
			\sqrt{ e^{-\lambda t}(\sigma_{i,0}^2 - \kappa_{i,1}) + \kappa_{i,1} }
			-
			\frac{\kappa_{i,2}(1 - e^{-2\lambda t})}
			{16\big( e^{-\lambda t}(\sigma_{i,0}^2 - \kappa_{i,1}) + \kappa_{i,1} \big)^{3/2}}, \quad i = 1,2,3,
		\end{align}
		whose absolute error is less than or equal to $\displaystyle \frac{\mu_3}{16(e^{-\lambda t}(\sigma_{i,0}^2 - \kappa_{i,1}) + \kappa_{i,1})^{5/2}}$ for each $i$.\\
		\\
		The expected value of equation \eqref{sigma^2_R} applying equations \eqref{E_Sigma_2}, \eqref{E0}, \eqref{E1}, \eqref{E2}, \eqref{E3}, \eqref{Var_i}, and $ \text{Var}\big[Z_{1}^{(0)}\big] = \kappa_{0,2} $ gives that
		\begin{align} \label{main}
			\EX^\Q \left[\sigma_{R}^2\right] 
			& = \frac{1}{T} \int_0^T \EX^\Q \left[\left| \Sigma_t^B \right|\right] dt \nonumber \\
			& = \frac{|C|}{T} \Big[ E_0
			+ \lambda \kappa_{0,2} \big(
			\delta_{11}\rho_1^2 E_1 + \delta_{22}\rho_2^2 E_2 + \delta_{33}\rho_3^2 E_3 \nonumber \\
			&  \qquad \qquad \qquad \qquad
			+ 2 \delta_{21}\rho_2\rho_1 E_4 + 2 \delta_{31}\rho_3\rho_1 E_5 + 2 \delta_{32}\rho_3\rho_2 E_6 \big) \Big],
		\end{align}	
		where  $ E_0 $, $ E_1 $, $ E_2 $, and $ E_3 $ are exactly given by
		\begin{align*}
			E_0 & = \int_0^T \prod_{i=1}^{3} \mathbb{E}^\Q[\sigma_{i,t}^2] dt \\
			& = \frac{1 - e^{-3\lambda T}}{3\lambda}
			(\sigma_{1,0}^2 - \kappa_{1,1})(\sigma_{2,0}^2 - \kappa_{2,1})(\sigma_{3,0}^2 - \kappa_{3,1}) \\
			&\quad 
			+ \frac{1 - e^{-2\lambda T}}{2\lambda}  \Big[
			\kappa_1^1 (\sigma_{2,0}^2 - \kappa_{2,1})(\sigma_{3,0}^2 - \kappa_{3,1}) 
			+ \kappa_1^2 (\sigma_{1,0}^2 - \kappa_{1,1})(\sigma_{3,0}^2 - \kappa_{3,1}) \\
			&\qquad \qquad \qquad \quad
			+ \kappa_1^3 (\sigma_{1,0}^2 - \kappa_{1,1})(\sigma_{2,0}^2 - \kappa_{2,1})
			\Big] \\
			&\quad 
			+ \frac{1 - e^{-\lambda T}}{\lambda}  \Big[
			\kappa_{1,1} \kappa_{2,1} (\sigma_{3,0}^2 - \kappa_{3,1}) 
			+ \kappa_{1,1} \kappa_{3,1} (\sigma_{2,0}^2 - \kappa_{2,1}) 
			+ \kappa_{2,1} \kappa_{3,1} (\sigma_{1,0}^2 - \kappa_{1,1})
			\Big] \\
			& \quad
			+ T \kappa_{1,1} \kappa_{2,1} \kappa_{3,1},
		\end{align*}
		
		\begin{align*}
			E_1 
			= & \int_0^T \mathbb{E}^\Q [\sigma_{3,t}^2] \mathbb{E}^\Q [\sigma_{2,t}^2 ] dt \\
			= &\ \frac{1 - e^{-2\lambda T}}{2\lambda}
			\big(\sigma_{3,0}^2 - \kappa_{3,1}\big)\big(\sigma_{2,0}^2 - \kappa_{2,1}\big)
			+ \frac{1 - e^{-\lambda T}}{\lambda}  \Big[
			\kappa_{3,1} (\sigma_{2,0}^2 - \kappa_{2,1})
			+ \kappa_{2,1} (\sigma_{3,0}^2 - \kappa_{3,1}) 
			\Big] \\
			& + T  \kappa_{3,1} \kappa_{2,1},
		\end{align*}
		
		\begin{align*}
			E_2 = & \int_0^T \mathbb{E}^\Q [\sigma_{3,t}^2] \mathbb{E}^\Q[\sigma_{1,t}^2] dt \\
			= &\ \frac{1 - e^{-2\lambda T}}{2\lambda}
			(\sigma_{3,0}^2 - \kappa_{3,1})(\sigma_{1,0}^2 - \kappa_{1,1})
			+ \frac{1 - e^{-\lambda T}}{\lambda} \Big[
			\kappa_{3,1} (\sigma_{1,0}^2 - \kappa_{1,1})
			+ \kappa_{1,1} (\sigma_{3,0}^2 - \kappa_{3,1}) 
			\Big] \\
			& + T \kappa_{3,1} \kappa_{1,1},
		\end{align*}
		
		\begin{align*}
			E_3 = & \int_0^T \mathbb{E}^\Q [\sigma_{2,t}^2] \mathbb{E}^\Q [\sigma_{1,t}^2] dt \\
			= &\ \frac{1 - e^{-2\lambda T}}{2\lambda} 
			(\sigma_{2,0}^2 - \kappa_{2,1})(\sigma_{1,0}^2 - \kappa_{1,1})
			+ \frac{1 - e^{-\lambda T}}{\lambda} \Big[
			\kappa_{2,1} (\sigma_{1,0}^2 - \kappa_{1,1})
			+ \kappa_{1,1} (\sigma_{2,0}^2 - \kappa_{2,1}) 
			\Big] \\
			& + T \kappa_{2,1} \kappa_{1,1},
		\end{align*}
		and $ E_4 $, $ E_5 $, and $ E_6 $ are approximated as
		\begin{align*}
			E_4 & = \int_0^T \EX^\Q[\sigma_{3,t}^2] \EX^\Q[\sigma_{2,t}] \EX^\Q[\sigma_{1,t}] dt \\
			& \approx 
			\int_0^T \Big[e^{-\lambda t}(\sigma_{3,0}^2 - \kappa_{3,1}) + \kappa_{3,1}\Big] \\
			& \qquad \quad
			\times \Bigg[\sqrt{ e^{-\lambda t}(\sigma_{2,0}^2 - \kappa_{2,1}) + \kappa_{2,1} }
			-
			\frac{\kappa_{2,2}(1 - e^{-2\lambda t})}
			{16\big( e^{-\lambda t}(\sigma_{2,0}^2 - \kappa_{2,1}) + \kappa_{2,1} \big)^{3/2}}\Bigg]\\
			& \qquad \quad 
			\times \Bigg[\sqrt{ e^{-\lambda t}(\sigma_{1,0}^2 - \kappa_{1,1}) + \kappa_{1,1} }
			-
			\frac{\kappa_{1,2}(1 - e^{-2\lambda t})}
			{16\big( e^{-\lambda t}(\sigma_{1,0}^2 - \kappa_{1,1}) + \kappa_{1,1} \big)^{3/2}}\Bigg]
			dt,
		\end{align*}
		
		\begin{align*}
			E_5 & = \int_0^T \EX^\Q[\sigma_{3,t}] \EX^\Q[\sigma_{2,t}^2] \EX^\Q[\sigma_{1,t}] dt \\
			& \approx 
			\int_0^T 
			\Bigg[\sqrt{ e^{-\lambda t}(\sigma_{3,0}^2 - \kappa_{3,1}) + \kappa_{3,1} }
			-
			\frac{\kappa_{3,2}(1 - e^{-2\lambda t})}
			{16\big( e^{-\lambda t}(\sigma_{3,0}^2 - \kappa_{3,1}) + \kappa_{3,1} \big)^{3/2}}\Bigg] \\
			& \qquad \quad 
			\times \Big[e^{-\lambda t}(\sigma_{2,0}^2 - \kappa_{2,1}) + \kappa_{2,1}\Big] \\
			& \qquad \quad 
			\times \Bigg[\sqrt{ e^{-\lambda t}(\sigma_{1,0}^2 - \kappa_{1,1}) + \kappa_{1,1} }
			-
			\frac{\kappa_{1,2}(1 - e^{-2\lambda t})}
			{16\big( e^{-\lambda t}(\sigma_{1,0}^2 - \kappa_{1,1}) + \kappa_{1,1} \big)^{3/2}}\Bigg]
			dt,
		\end{align*}
		
		\begin{align*}
			E_6 & = \int_0^T \EX^\Q[\sigma_{3,t}] \EX^\Q[\sigma_{2,t}] \EX^\Q[\sigma_{1,t}^2] dt \\
			& \approx 
			\int_0^T
			\Bigg[\sqrt{ e^{-\lambda t}(\sigma_{3,0}^2 - \kappa_{3,1}) + \kappa_{3,1} }
			-
			\frac{\kappa_{3,2}(1 - e^{-2\lambda t})}
			{16\big( e^{-\lambda t}(\sigma_{3,0}^2 - \kappa_{3,1}\big) + \kappa_{3,1})^{3/2}}\Bigg] \\
			& \qquad \quad 
			\times \Bigg[\sqrt{ e^{-\lambda t}(\sigma_{2,0}^2 - \kappa_{2,1}) + \kappa_{2,1} }
			-
			\frac{\kappa_{2,2}(1 - e^{-2\lambda t})}
			{16\big( e^{-\lambda t}(\sigma_{2,0}^2 - \kappa_{2,1}) + \kappa_{2,1} \big)^{3/2}}\Bigg] \\
			& \qquad \quad  
			\times \Big[e^{-\lambda t}(\sigma_{1,0}^2 - \kappa_{1,1}) + \kappa_{1,1}\Big]
			dt.
		\end{align*}			
		Finally, the theorem uses equation \eqref{main} to complete the proof.
	\end{proof}

	\section{Model Fitting and Parameter Estimation}
	\label{sec5}
	
	This section presents numerical results, model calibration, and parameter testing procedures. For the purpose of model calibration, nine stocks were selected using the ``quantmod" package (\cite{quantmod}) in R \cite{RRR}, covering the period from January 1, 2020, to January 1, 2024. The mean, variance, and kurtosis of the 9 assets can be also found in appendix \ref{Summaryb}. These stocks were then randomly reshuffled into three groups, each consisting of three stocks, to evaluate the robustness and precision of the proposed model in different combinations of assets. The log-returns of the daily closing prices were computed for each stock. Subsequently, the covariance matrix of the log-returns was estimated at 10-day time intervals, and the determinant of each covariance matrix was calculated to capture the joint variability and dependency structure among the selected assets over time.  The correlation matrix of Coca-Cola, Apple, and Tesla, the correlation matrix for Google, Microsoft, and Meta, and the correlation matrix for J.P. Morgan, Nvidia, and Amazon are in Figure \ref{fig:correlation}.

	\begin{figure}[H]
		\centering
		
		\begin{subfigure}[b]{0.32\textwidth}
			\includegraphics[width=\textwidth]{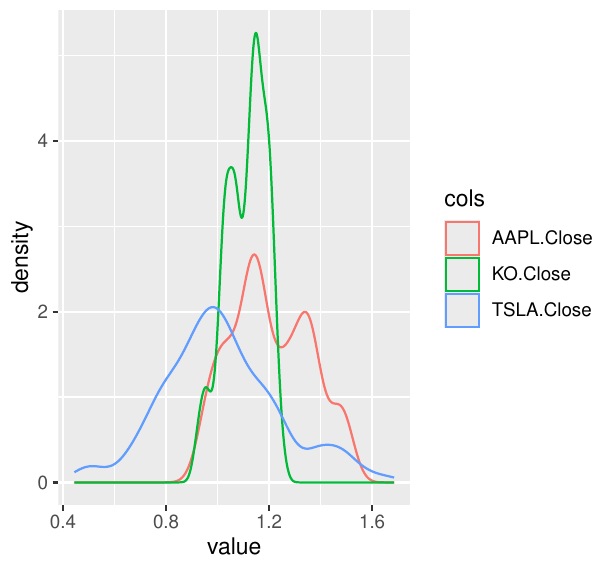}
			\caption{KO, AAPL,TSLA}
		\end{subfigure}
		\hfill
		\begin{subfigure}[b]{0.32\textwidth}
			\includegraphics[width=\textwidth]{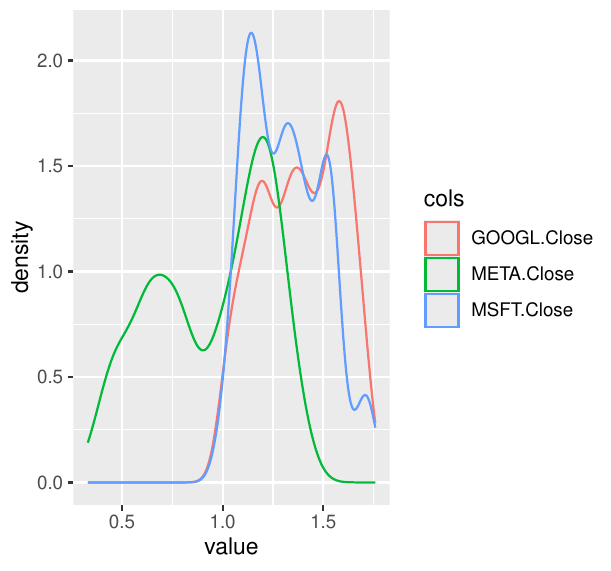}
			\caption{GOOGL, MSFT, META,}
		\end{subfigure}
		\hfill
		\begin{subfigure}[b]{0.32\textwidth}
			\includegraphics[width=\textwidth]{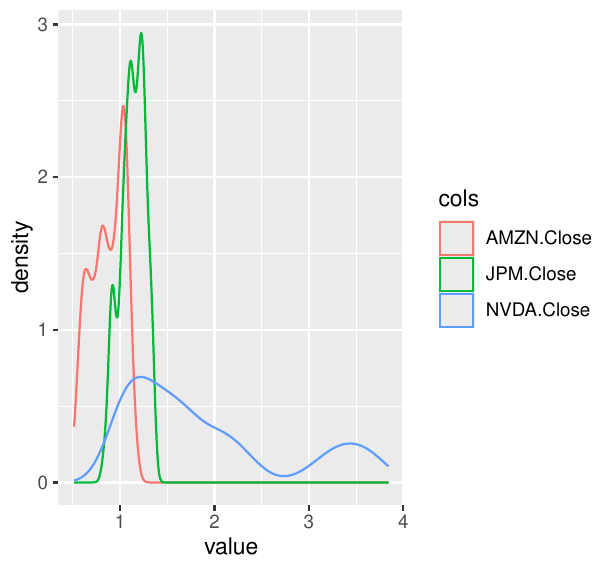}
			\caption{JPM, NVDA, AMZN}
		\end{subfigure}
		
		\caption{Grouped line histogram of 9 stocks over the period 2021–2024.}
		\label{fig1}
	\end{figure}
	Figure \ref{fig1} presents line histograms of the log-return distributions of the three stocks grouped together for analysis. These histograms reveal that the underlying assets exhibit similar distributional characteristics and are therefore grouped together for further analysis in the subsequent section of this paper. In particular, the study focuses on assets that share comparable statistical properties, including mean, variance, kurtosis, and skewness. The observed similarities in the log-return distributions suggest that these assets may respond to market factors in a relatively consistent manner. Such consistency provides a reasonable and statistically justified basis for conducting a joint variance swap analysis. Moreover, grouping assets with similar return dynamics allows for a more coherent comparison of volatility behavior and improves the reliability of the empirical analysis presented in this study.

	\begin{figure}[H]
		\centering
		
		\begin{subfigure}[b]{0.32\textwidth}
			\includegraphics[width=\textwidth]{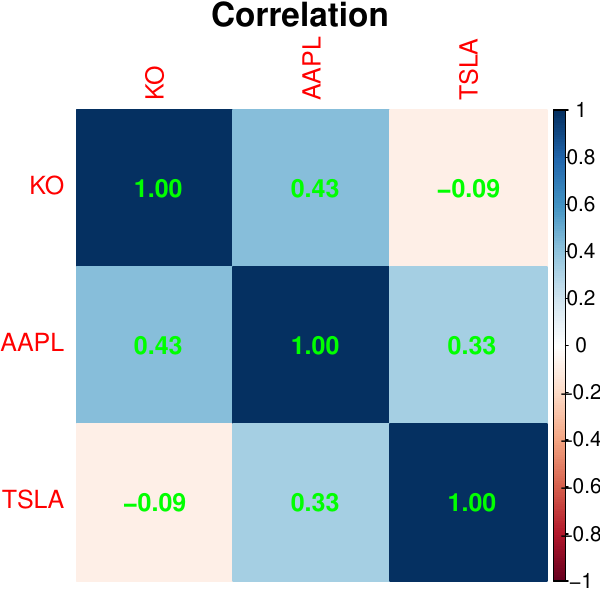}
			\caption{KO, AAPL,TSLA}
		\end{subfigure}
		\hfill
		\begin{subfigure}[b]{0.32\textwidth}
			\includegraphics[width=\textwidth]{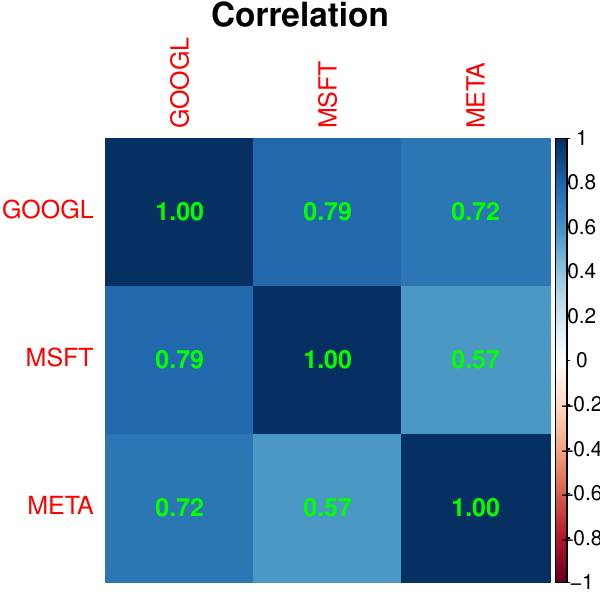}
			\caption{GOOGL, MSFT, META,}
		\end{subfigure}
		\hfill
		\begin{subfigure}[b]{0.32\textwidth}
			\includegraphics[width=\textwidth]{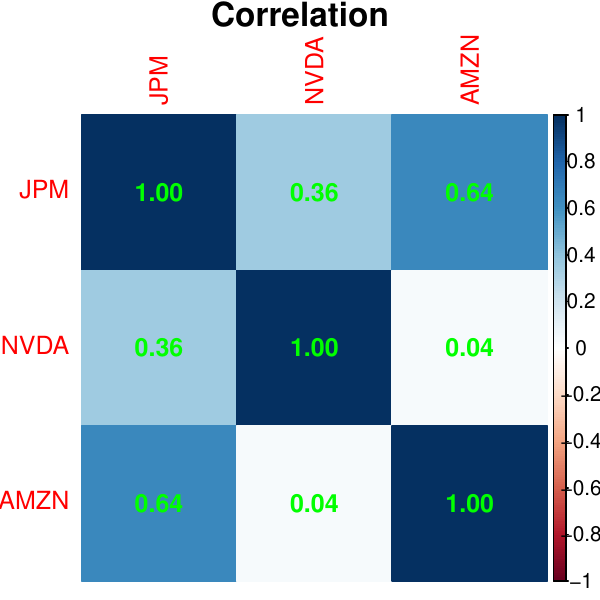}
			\caption{JPM, NVDA, AMZN}
		\end{subfigure}
		
		\caption{Grouped correlations of 9 stocks over the period 2021–2024.}
		\label{fig:correlation}
	\end{figure}

	Figure \ref{fig:correlation} illustrates the correlation matrix among the three underlying assets considered in this analysis. The correlation heatmap indicates that the assets are generally moderately to weakly correlated with one another, except for META and Google, which exhibit a relatively strong positive correlation with a coefficient of 0.72. This suggests that META and Google tend to move in similar directions more frequently compared to the other asset pairs.  
	\begin{figure}[H]
		\centering
		
		\begin{subfigure}[b]{0.32\textwidth}
			\includegraphics[width=\textwidth]{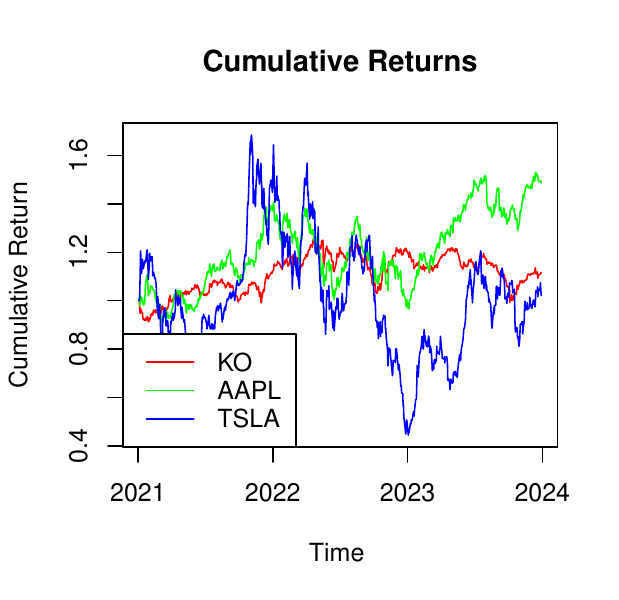}
			\caption{KO, AAPL,TSLA}
			\label{fig:2a}
		\end{subfigure}
		\hfill
		\begin{subfigure}[b]{0.32\textwidth}
			\includegraphics[width=\textwidth]{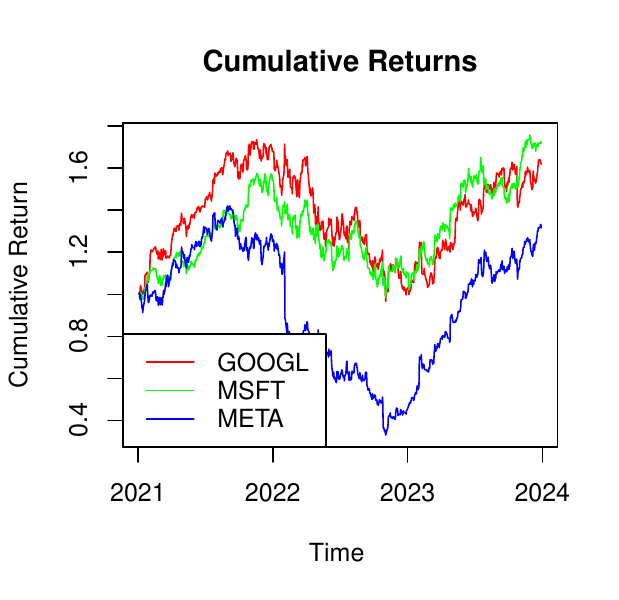}
			\caption{GOOGL, MSFT, META,}
			\label{fig:2b}
		\end{subfigure}
		\hfill
		\begin{subfigure}[b]{0.32\textwidth}
			\includegraphics[width=\textwidth]{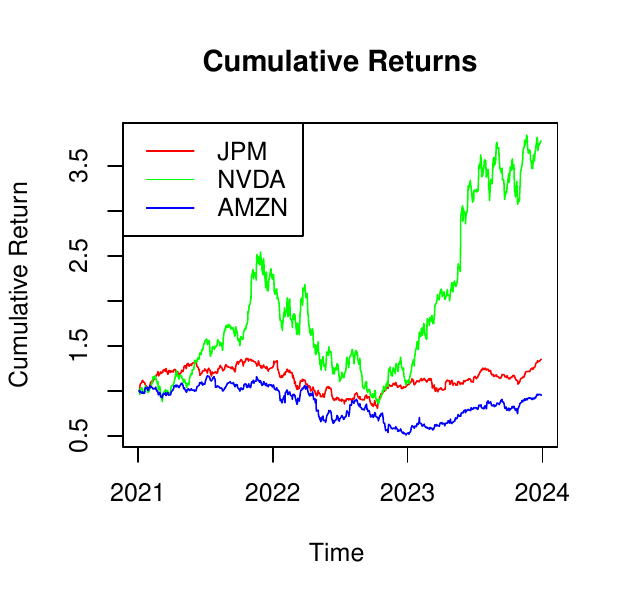}
			\caption{JPM, NVDA, AMZN}
			\label{fig:2c}
		\end{subfigure}
		
		\caption{Grouped cumulative returns of 9 stocks over the period 2021–2024.}
		\label{fig:cumulative_returns}
	\end{figure}
	
	Figure \ref{fig:cumulative_returns} illustrates the cumulative returns of the closing prices for the nine selected assets, grouped as follows: the left panel presents Coca-Cola, Apple, and Tesla; the middle panel displays Google, Microsoft, and Meta; and the right panel includes JPMorgan, Nvidia, and Amazon. The figure clearly shows that the cumulative return distributions of these assets vary substantially. Some assets, such as Tesla and Nvidia, exhibit high volatility, whereas others, like J.P. Morgan, demonstrate relatively stable performance. The rationale for selecting this diverse set of covariates is to evaluate the robustness and generalization capability of the model’s predictive accuracy across assets with different volatility and market behaviors. 
	
	\begin{figure}[H]
		\centering
		
		\begin{subfigure}[b]{0.32\textwidth}
			\includegraphics[width=\textwidth]{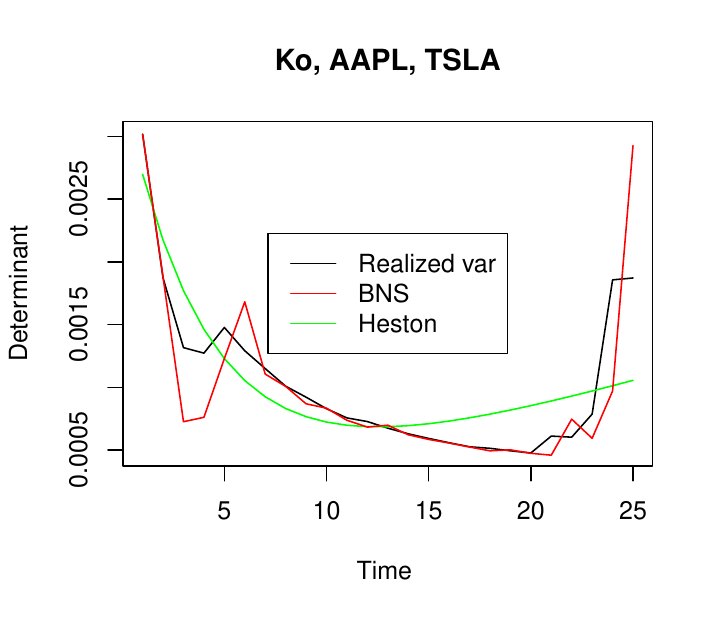}
			\caption{KO, AAPL, TSLA}
			\label{fig:3a}
		\end{subfigure}
		\hfill
		\begin{subfigure}[b]{0.32\textwidth}
			\includegraphics[width=\textwidth]{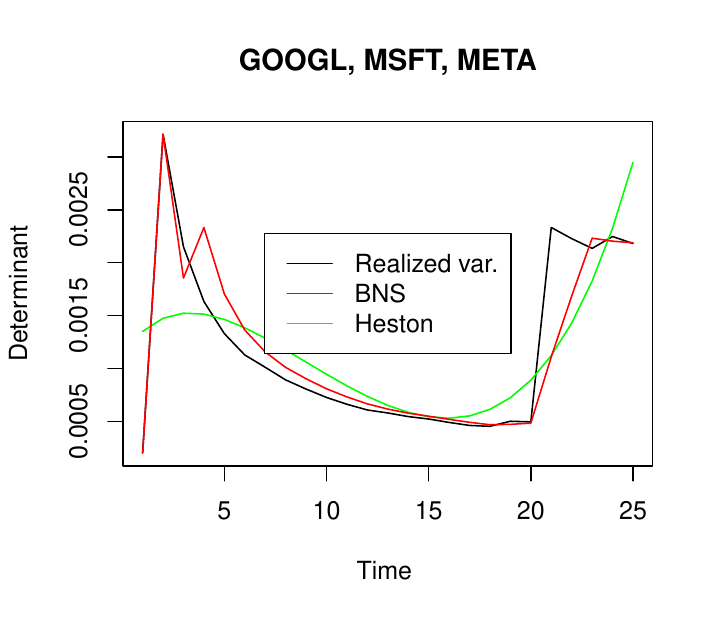}
			\caption{GOOGL, MSFT, META}
			\label{fig:3b}
		\end{subfigure}
		\hfill
		\begin{subfigure}[b]{0.32\textwidth}
			\includegraphics[width=\textwidth]{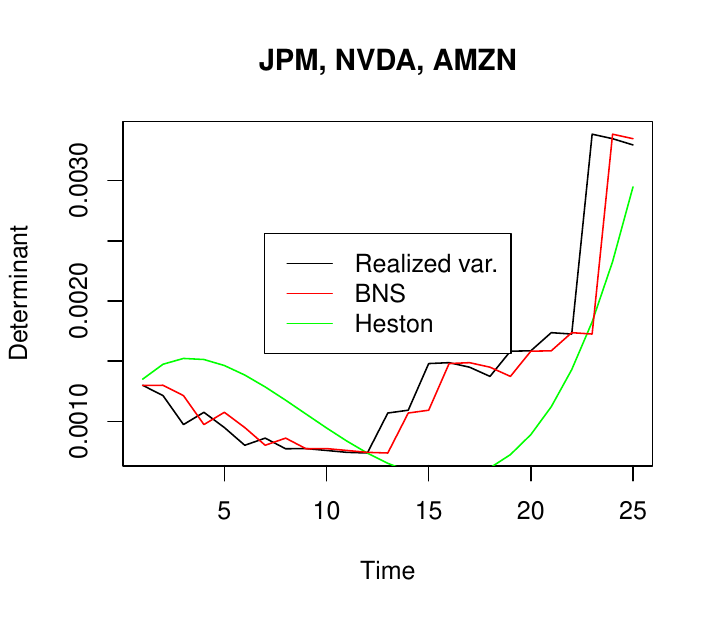}
			\caption{JPM, NDVA, AMZN}
			\label{fig:3c}
		\end{subfigure}
		
		\caption{Realized vs Fitted Generalized Dispersion Swap under Heston and BNS Models}
		\label{fig:Realised variance}
	\end{figure}
	
	Figure \ref{fig:Realised variance} presents the realized variance of the covariance matrix along with the fitted results obtained from the Heston and BNS models. A nonlinear least squares (NLS) estimator was employed in R for parameter calibration and model validation. The estimation results indicate that all parameters in both the Heston and BNS models were statistically significant, with $p$-values on the order of $2.6 \times10^{-16}$, confirming the overall robustness and suitability of the fitted models.\\
	\\
	Based on these findings, both the Heston and BNS models demonstrated strong performance in capturing and predicting the determinant of the realized variance. However, the BNS model outperformed the Heston model in several key aspect it achieved higher predictive accuracy, reduced the mean squared error, and exhibited faster convergence. Quantitatively, the BNS model was able to explain approximately 40\% more of the variation in the data that the Heston model fails to capture. Hence, on average, the BNS model provided about a 40\% improvement in performance over the standard Heston model.

	\begin{table}[H]
		\caption{Parameter estimates of KO, APPL, and TSLA using Heston Model}
		\label{Heston1Ko}
		\centering
		\begin{tabular}{|c||c| c| r|r| }      
			\hline\hline 
			Parameters&Estimate & Standard Error &t Value& $Pr(>|t|) $\\[0.5ex]
			\hline
			$\kappa_1$ & 0.2085 & 0.0461&4.52  &$ 0.0002$ \\
			$\kappa_2$ & 0.2530 & 0.0060&42.17&$<0.0001$ \\
			$\kappa_3$ & 0.0413 & 0.0026&15.88  &$ 0.0005$ \\
			$\theta_1^2$& 0.2159& 0.0812&2.66 &$<0.0001 $ \\
			$\theta_2^2$& 0.2102& 0.0636&3.31&$<0.0001 $ \\
			$\theta_3^2$& 0.1837& 0.0157&11.70 &$<0.0001 $\\[1ex]     
			\hline
		\end{tabular}
	\end{table}
	
	\begin{table}[H]
		\caption{Parameter estimates of GOOG, MSFT, and META using Heston Model}
		\label{Heston2}
		\centering
		\begin{tabular}{|c||c| c| r|r| }     
			\hline\hline 
			Parameters&Estimate & Standard Error &t Value& $Pr(>|t|) $\\[0.5ex]
			\hline
			$\kappa_1$& 0.1337 & 0.0457&2.93&$0.0081$ \\
			$\kappa_2$ &0.0182& 0.0050&3.63&$0.0016$ \\
			$\kappa_3$ & 0.2013& 0.0015&134.20&$<0.0001$ \\
			$\theta_1^2$& 0.1589&0.0276&5.76&$<0.0001$ \\
			$\theta_2^2$& 0.0175&0.0024&7.29&$0.0006 $ \\
			$\theta_3^2$&0.1644&0.0018&91.33&$<0.0001 $\\[1ex]   
			\hline
		\end{tabular}
	\end{table}

	\begin{table}[H]
		\caption{Parameter estimates  of JPM, NVIDA, and AMZN using Heston Model}
		\label{Heston3}
		\centering
		\begin{tabular}{|c||c| c| r|r| }   
			\hline\hline 
			Parameters&Estimate & Standard Error &t Value& $Pr(>|t|) $\\[0.5ex]
			\hline
			$\kappa_1$& 0.0236 & 0.0291&0.81&$0.0215$ \\
			$\kappa_2$ &0.3586& 0.0700&5.13&$<0.0001$ \\
			$\kappa_3$ & 0.1997& 0.0002&998.50&$<0.0001$ \\
			$\theta_1^2$& 0.0389&0.0032&12.15&$<0.0001$ \\
			$\theta_2^2$& 0.1226&0.0513&2.39&$<0.0001$ \\
			$\theta_3^2$&0.0128&0.00682&1.88&$<0.0001$\\[1ex]    
			\hline
		\end{tabular}
	\end{table}

	\begin{table}[H]
		\caption{Parameter estimates of KO, APPL, and TSLA using BNS Model}
		\label{BNS1}
		\centering
		\begin{tabular}{|c||c| c| r|r| }    
			\hline\hline 
			Parameters&Estimate & Standard Error &t Value& $Pr(>|t|) $\\[0.5ex]
			\hline
			$\lambda$& 1.5610 & $<0.0001$&178.34&$<0.0001$ \\
			$\kappa_1$& 0.0134 & 0.0012&18.46&$<0.0001$ \\
			$\kappa_2$ &0.5891& 0.0002&29.14&$<0.0001$ \\
			$\kappa_3$ & 0.2732& 0.0007&104.23&$<0.0001$ \\
			$\theta_1^2$& 0.0386&0.0136&89.46&$<0.0001 $ \\
			$\theta_2^2$& 0.0457&0.0001&116.79&$<0.0001 $ \\
			$\theta_3^2$&0.8975&0.0012&234.12&$<0.0001 $\\[1ex]     
			\hline
		\end{tabular}
	\end{table}
	
	\begin{table}[H]
		\caption{Parameter estimates of GOOG, MSFT, and META using BNS Model}
		\label{BNS2}
		\centering
		\begin{tabular}{|c||c| c| r|r| }       
			\hline\hline 
			Parameters&Estimate & Standard Error &t Value& $Pr(>|t|) $\\[0.5ex]
			\hline
			$\lambda$& 0.1897 & 0.0011&172.45&$<0.0001$ \\
			$\kappa_1$& 0.1201 & 0.0139&8.64&$0.0005$ \\
			$\kappa_2$ &0.8912& 0.014&63.66&$<0.0001$ \\
			$\kappa_3$ & 0.1121& 0.0013&86.23&$<0.0001$ \\
			$\theta_1^2$& 0.4571&0.0024&190.45&$<0.0001$ \\
			$\theta_2^2$& 0.0139&0.0112&1.24&$0.0001 $ \\
			$\theta_3^2$&0.7813&0.08711&9.33&$<0.0001$\\[1ex]         
			\hline
		\end{tabular}
	\end{table}
	
	\begin{table}[H]
		\caption{Parameter estimates of JPM, NVIDA, and META using BNS Model}
		\label{BNS3}
		\centering
		\begin{tabular}{|c||c| c| r|r| }       
			\hline\hline 
			Parameters&Estimate & Standard Error &t Value& $Pr(>|t|) $\\[0.5ex]
			\hline
			$\lambda$& 0.0231 & 0.0145&1.59.78&$<0.0001$ \\
			$\kappa_1$& 0.0130 & 0.0014&9.28&0.0011 \\
			$\kappa_2$ &0.0314& 0.0003&104.66&$<0.0001$ \\
			$\kappa_3$ & 0.0146& 0.0130&61.5&$<0.0001$ \\
			$\theta_1^2$& 0.0184&0.0008&23.00&$<0.0001$ \\
			$\theta_2^2$& 0.0146&0.0022&6.63&0.0013 \\
			$\theta_3^2$&0.1351&0.0130&10.39&$<0.0001$\\[1ex]         
			\hline
		\end{tabular}
	\end{table}
	The above tables (\ref{Heston1Ko}–\ref{BNS3}) present the parameter estimates together with the corresponding p-values obtained from the t-tests used to assess the statistical significance of each parameter. As shown in these tables, all estimated parameters are statistically significant at conventional significance levels, indicating that they are significantly different from zero. This suggests that each parameter plays an important role in the model specification and contributes meaningfully to the prediction and modeling of generalized multivariate portfolios under both the Heston and Barndorff-Nielsen and Shephard (BNS) stochastic volatility frameworks. 
	\par
	
	Multiple error metrics are employed to evaluate the performance and accuracy of our models, including the Absolute Percentage Error (APE), Average Absolute Error (AAE), Average Relative Percentage Error (ARPE), and Root Mean Square Error (RMSE). These metrics provide a comprehensive assessment of model goodness-of-fit by capturing different aspects of prediction error both in magnitude and variability. The formal definitions of these error measures, as summarized in Tables \ref{errs} and \ref{errsBNS}.
	\begin{table}[H]
		\centering
		\caption{Error measurement when Heston model is implemented.}
		\label{errs}
		\begin{tabular}{|c||r| r| r|r| }       
			\hline
			Covariate&RMSE &APE &AAE&ARPE\\[0.5ex]
			\hline
			
			KO, AAPL, TSLA&0.2528& 0.0003 &\textless0.0001& 0.0003\\
			GOOGL, MSFT, META&0.3231& 0.0004& \textless0.0001 &0.0006\\
			JPM, NVDA, AMZN&5.1830& 0.0003& \textless0.0001& 0.0003\\     
			\hline
		\end{tabular}
	\end{table}
	
	\begin{table}[H]
		\centering
		\caption{Error measurement when BNS model is implemented.}
		\label{errsBNS}
		\begin{tabular}{|c||r| r| r|r| }       
			\hline
			Covariate&RMSE &APE &AAE&ARPE\\[0.5ex]
			\hline
			
			KO, AAPL, TSLA& 0.1708& 0.0002& \textless0.0001 &0.0001\\
			GOOGL, MSFT, META& 0.1704&\textless0.0001&\textless0.0001&0.0001\\
			JPM, NVDA, AMZN&1.1088 &0.0002& \textless0.0001 &0.0002\\     
			\hline
		\end{tabular}
	\end{table}
	
	Tables  \ref{errs} and \ref{errsBNS}  present the numerical error calculations for the realized variance of the multi-asset determinant covariate matrices. The reported error metrics provide a quantitative comparison between the models, highlighting their relative predictive performance. As shown in these tables, the BNS model consistently demonstrates lower error values across all measures, indicating a superior fit and improved accuracy compared to the standard Heston model. This result suggests that the BNS framework captures the underlying market dynamics more effectively, particularly in modeling volatility and cross-asset relationships. 
	
	\section{Conclusion}
	\label{sec6}

	In this paper, we introduced a generalized variance method for computing the determinant-based instantaneous generalized variance of multi-asset log-returns under the Heston and Barndorff-Nielsen–Shephard (BNS) models. By exploiting the determinant of the multi-asset return covariance matrix, the proposed method provides a tractable framework for estimating generalized realized variance and evaluating diversified multi-asset generalized dispersion swap. This determinant-based measure captures the joint dispersion of asset returns and accounts for their dependence structure.
	
	We applied the proposed method to nine selected stocks, which were randomly partitioned into three groups of three distinct stocks. The fitted Heston and BNS models both demonstrated strong agreement with the empirical generalized realized variance across the selected time horizons. However, the BNS model provided a superior fit, explaining approximately 40\% more of the variation in the data that remained unexplained by the Heston model. The BNS model also yielded lower estimation errors, indicating improved empirical accuracy.
	
	These findings are consistent with the ability of the BNS framework to incorporate jumps in the underlying asset dynamics, allowing it to capture features of financial returns that may not be adequately represented by the continuous diffusion component of the Heston model. Overall, the results highlight the potential of the proposed generalized variance method as a practical tool for modeling multi-asset dependence and evaluating determinant-based generalized dispersion swap.

	\begin{appendices}
		\section{Calculation of Determinants} 
		\label{determinanta}
		
		If the 3-dimensional covariance matrix for Heston model is given by \eqref{eq:heston_three_covariance_matrix}, it is trivial that $ \sigma_{i,t}^2 = \sigma_{i,t} \sigma_{i,t} $, $ i = 1, 2, 3 $. Then
		\begin{align}\label{DCD}
			\Sigma_t^H = D_t^H C D_t^H,
		\end{align}
		where $ C = (c_{ij})_{1\leq i,j \leq 3} $ is the correlation matrix of stock prices which can be calculated using stock price data, and $ D_t^H = \text{diag}(\sigma_{1,t}, \sigma_{2,t}, \sigma_{3,t}) $.
		Hence, $\left|\Sigma_t^H\right| = |C| \left|D_t^H\right|^2$ gives equation \eqref{eq:hes21}.\\
		\\
		If the 3-dimensional covariance matrix for BNS model is given by \eqref{eq:bns_three_stock_continuous_covariance}, then
		\begin{align*}
			\widetilde{\Sigma}_t^{B} = \Sigma_t^{B,c} + \lambda \kappa_{0,2} {\rho} {\rho}^\top,
		\end{align*}
		where $\rho=(\rho_1,\rho_2,\rho_3)^\top$.
		By the matrix determinant lemma,
		\begin{align}\label{Sigma_2}
			\left| \widetilde{\Sigma}_t^{B} \right| 
			= \left| \Sigma_t^{B,c} + (\lambda \kappa_{0,2} {\rho}) {\rho}^\top \right|
			= \left| \Sigma_t^{B,c} \right| \left(1 + \lambda \kappa_{0,2} {\rho}^\top \left(\Sigma_t^{B,c}\right)^{-1}  {\rho} \right).
		\end{align}
		We notice that, at fixed time $t$, the correlation matrix $ C(t) $ of stock prices can be calculated using stock price data and be regarded as a constant matrix $ C$.
		Simultaneously, $ C^{-1} $ can be also calculated using the same stock price data.
		Therefore, we denote $ C^{-1} = (\delta_{ij})_{1\leq i,j \leq 3} $.
		By equation \eqref{DCD}, we have
		\begin{align}\label{Sigma-1}
			\left( \Sigma_t^{B,c} \right)^{-1}
			= \left( D_t^B\right )^{-1} \left( R^B \right)^{-1} \left( D_t^B \right)^{-1}
			= \begin{pmatrix}
				\frac{\delta_{11}}{\sigma_{1,t}^2} & \frac{\delta_{12}}{\sigma_{1,t}\sigma_{2,t}} & \frac{\delta_{13}}{\sigma_{1,t}\sigma_{3,t}}\\
				\frac{\delta_{21}}{\sigma_{2,t}\sigma_{1,t}} & \frac{\delta_{22}}{\sigma_{2,t}^2} & \frac{\delta_{23}}{\sigma_{2,t}\sigma_{3,t}}\\
				\frac{\delta_{31}}{\sigma_{3,t}\sigma_{1,t}} & \frac{\delta_{32}}{\sigma_{3,t}\sigma_{2,t}} & \frac{\delta_{33}}{\sigma_{3,t}^2}
			\end{pmatrix}.
		\end{align}
		Finally, we obtain $\left| \widetilde{\Sigma}_t^{B} \right|$ in equation \eqref{Sigma_2_det} using \eqref{eq:hes21}, \eqref{Sigma_2}, and \eqref{Sigma-1}.

		\section{Summary Statistics } 
		\label{Summaryb}
		
		\begin{table}[H]
			\centering
			\caption{Summary statistics of the nine assets.}
			
			\begin{tabular}{|l||c| c| c|r| }       
				\hline
				Closing Price&Mean &Variance &Kurtosis\\[0.5ex]
				\hline
				
				Ko.Close    &0.1095    &     0.0061   &       -0.6261\\
				APPL.Close  &  1.2061      &   0.0236 &        -0.9089\\
				TSLA.Close  & 1.0150       &  0.0522  &         0.1955\\
				GOOGL.Close &1.3819        & 0.0403   &       -1.1390\\
				MSFT.Close&1.3148 &0.0342    &      -0.8078\\
				META.Close&0.9457      &   0.0849     &     -1.1963\\
				JPM.Close&1.1330       &  0.0162      &    -0.6844\\
				NVDA.Close&  1.8961      &   0.7287      &    -0.4353\\
				AMZN.Close &  0.8679        & 0.0297     &    -1.1495\\   
				\hline
			\end{tabular}
		\end{table}

\end{appendices}

	\end{document}